\documentclass[aps,prl,twocolumn,superscriptaddress]{revtex4-1}

\usepackage{amsmath,amssymb}
\usepackage{mathrsfs}
\usepackage{graphicx}
\usepackage{color}
\usepackage{soul}
\usepackage[hidelinks]{hyperref}

\hypersetup{colorlinks=true, linkcolor=black, citecolor=black, urlcolor=blue}

\newcommand{\kgc}[1]{{\color{black} {#1}}}
\newcommand{\tb}[1]{{\color{black} {#1}}}

\begin{document}

\author{Karol Gietka}
\email[Corresponding author: ]{karol.gietka@oist.jp}
\author{Ayaka Usui}
\author{Jianqiao Deng}
\author{Thomas Busch}
\affiliation{Quantum Systems Unit, Okinawa Institute of Science and Technology Graduate University, Onna, Okinawa 904-0495, Japan} 

\title{Simulating the same physics with two distinct Hamiltonians}


\begin{abstract}
\tb{We develop a framework and give an example for situations where two}
\kgc{distinct Hamiltonians living in the same Hilbert space} 
can be used 
to simulate the \tb{same} \emph{physics}. 
As an example \kgc{of an analog simulation}, we first discuss how one can simulate an infinite-range-interaction one-axis twisting Hamiltonian using a short-range nearest-neighbor-interaction Heisenberg XXX model with a staggered field. Based on this, we show how one can  build an alternative version \kgc{of a digital quantum simulator}. 
As a by-product, we present a method for creating many-body \kgc{maximally} entangled states using only short-range nearest-neighbor interactions.  
\end{abstract}
\date{\today}
\maketitle

\emph{Introduction}---The concept behind quantum simulators is fairly straightforward to understand \cite{RevModPhys.86.153,whatis}, however, extremely challenging from an experimental point of view \cite{cirac2012goals}. Imagine that one has a target Hamiltonian $\hat H_\mathrm{T}$, and wants to study its properties or the dynamics governed by it. 
However, the system is either too large to perform numerical and analytical calculations on or is intractable from an experimental point of view. In this case, one can either come up with some other physical system that has a Hamiltonian $\hat H_\mathrm{QS}$ that is identical to $\hat H_{\mathrm{T}}$ and therefore possesses the same system properties and leads to the same dynamics. Or one can perform the desired evolution based on $\hat H_{\mathrm{T}}$ using an approximative stroboscopic time evolution through \emph{quantum kicks}.  The first case describes so-called analog quantum simulators, and the second case digital quantum simulators \cite{lloyd1996universal}. The idea of quantum simulators is commonly attributed to Richard Feynman who proposed it in 1982~\cite{feynman1982simulating}; however due to the experimental difficulties, in particular, controlling and tuning Hamiltonian parameters with high fidelities, the first viable ideas for quantum simulators were only proposed and realized very recently \cite{greiner2002quantum, lewenstein2007ultracold, lanyon2011universal, jotzu2014experimental, weimer2010rydberg,barreiro2011open, bernien2017probing, smith2016many, zhang2017observation, kim2010quantum, simon2011quantum, islam2011onset, britton2012engineered, muniz2020exploring} on a number of experimental platforms including ultra-cold quantum gases \cite{bloch2012quantum,gross2017quantum}, trapped ions \cite{blatt2012quantum}, photonic systems \cite{aspuru2012photonic}, and superconducting circuits \cite{houck2012chip}. 

The requirement for $\hat H_\mathrm{QS}$ to be a suitable Hamiltonian of a quantum simulator can be formulated in the following way (for the sake of brevity, we set $\hbar = 1$ throughout the entire manuscript)
\begin{align}\label{eq:condgen}
	  \langle \psi| e^{i t \hat H_\mathrm{QS}}e^{-i t \hat H_\mathrm{T}}| \psi \rangle = e^{i  \xi(t)},
\end{align}
where $\xi(t)$ is a \tb{{mostly real-valued}} function of time. \kgc{If the imaginary part of $\xi$ is zero, i.e., $\Im({\xi}) = 0$ then $\hat H_\mathrm{QS}$ is an ideal simulator, whereas if $\Im({\xi}) \neq 0$ the simulator is only suitable for times during which $\Im[{\xi(t)}] \ll 1$}. \kgc{In the ideal case,} the original idea of a quantum simulator considered either $\xi=0$ so that $ \hat H_\mathrm{QS} = \hat H_\mathrm{T}$, or $\xi$ to be some real-valued number $c$ multiplied by time, so that $\hat H_\mathrm{QS} = \hat H_\mathrm{T} + c\hat I$, where $\hat I$ is the identity operator. Making use of the Baker-Campbell-Hausdorff formula it is straightforward to write
\begin{align}
	  \langle \psi| e^{i  \hat h(t)}| \psi \rangle = e^{i  \xi(t)},
\end{align}
where $\hat h(t)$ is some, in general, time-dependent hermitian operator
\begin{align}\label{eq:BCH}
	\hat h(t) = t\left(\hat H_{\mathrm{QS}} -\hat H_{\mathrm{T}} \right)+\frac{i t^2}{2}\left[\hat H_{\mathrm{QS}}, -\hat H_{\mathrm{T}}   \right]+\ldots,
\end{align}
where $\left[\bullet,\bullet \right]$ stands for the commutator and $\ldots$ indicates terms involving higher order commutators of $\hat H_{\mathrm{QS}}$ and $\hat H_{\mathrm{T}}$.

\begin{figure}[tb]
\centering
    \includegraphics[width=0.47\textwidth, height=0.14\textwidth]{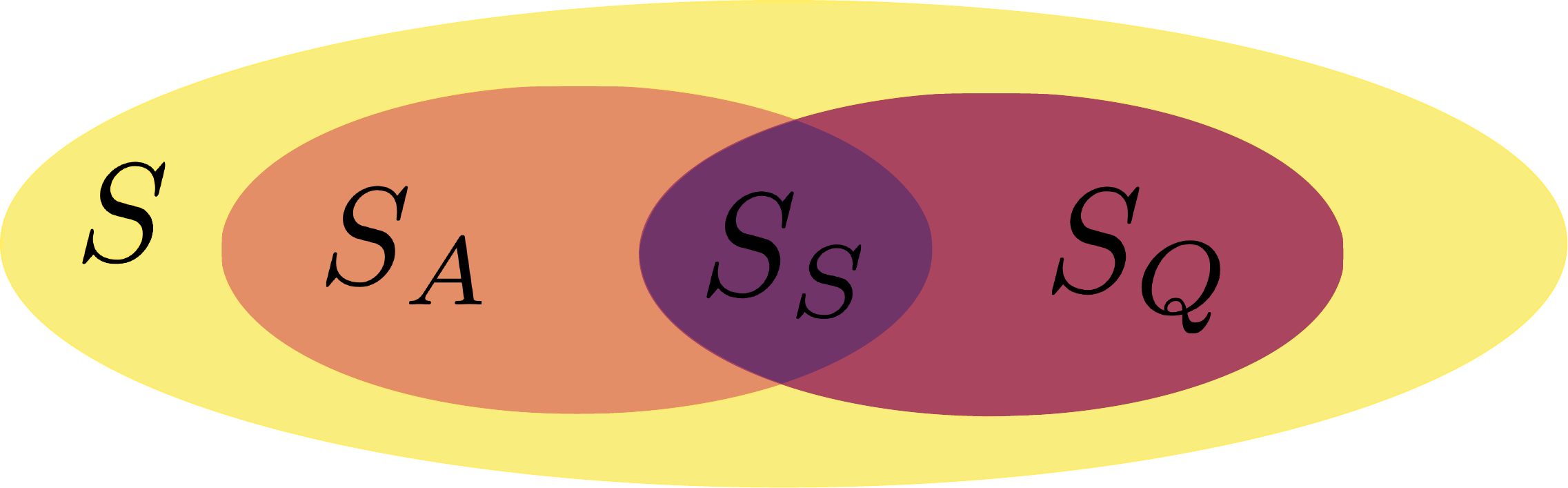}
 \caption{The set of all initial quantum states is given by $S$. Even if a quantum simulator Hamiltonian $\hat H_{\mathrm{QS}}$ is not the same as the target Hamiltonian $\hat H_{\mathrm{T}}$, there still exists some set of states $S_Q$ that can simulate the physics of $\hat H_{\mathrm{T}}$. If the overlap $S_S$ between the set of experimentally \kgc{accessible} states $S_A$ and $S_Q$ is not zero, the concept of using  the \kgc{connector} can be used in analog quantum simulators. The same concept can be also extended to the digital version of a quantum simulator by applying \emph{quantum kicks} using $\hat H_{\mathrm{QS}}$. The size of $S_Q$ depends on the form of $\hat H_{\mathrm{QS}}$. If $\hat H_{\mathrm{QS}} = \hat H_{\mathrm{T}}$, then $S = S_Q$ and $S_A = S_S$ is always a subset of $S_Q$.}
    \label{fig:scheme}
\end{figure}

The interpretation of $\hat h(t)$ is then straightforward. It is nothing else but an operator \kgc{which transforms  dynamics governed by $\hat H_\mathrm{T}$ to dynamics governed by $\hat H_\mathrm{QS}$}. If one knows the $\hat h(t)$ that relates the two Hamiltonians $\hat H_\mathrm{T}$ and $\hat H_\mathrm{QS}$, it is possible to simulate dynamics generated by $\hat H_\mathrm{T}$ by $\hat H_\mathrm{QS}$ by using the transformation
\begin{align}
	\langle \hat O (t) \rangle_{\mathrm{T}} = \langle e^{-i \hat h(t)} \hat O (t) e^{i \hat h (t)}\rangle_{\mathrm{QS}},
\end{align}
for any observable $\hat O (t)$. \kgc{For brevity, we will call $\hat h(t)$ a \kgc{connector} operator or simply connector.}
Unfortunately, due to their construction, connectors are likely to be rather complicated, time-dependent, or even non-local, and therefore most often of no practical help. However, we will show in the following that under certain conditions one can connect the dynamics governed by two \emph{substantially} different Hamiltonians \tb{in a valuable way}. One such condition is given in situations where two Hamiltonians commute, i.e., $\hat h(t) = \hat h \cdot t$. \kgc{This means that $\hat h$, $\hat H_{\mathrm{QS}}$, and $\hat H_{\mathrm{T}}$ share the same eigenbasis but have different eigenspectra. Then, if $\hat h$ has a degenerate eigenspectrum, it might happen that a state $|\psi \rangle$ composed of degenerate eigenstates of $\hat h = \hat H_{\mathrm{QS}} -\hat H_{\mathrm{T}}$ will not be an eigenstate of $\hat H_{\mathrm{QS}}$ or $\hat H_{\mathrm{T}}$,\tb{but} the two Hamiltonians will yield the same quantum dynamics with respect to that state (see Fig.~\ref{fig:scheme})}. Of course, finding two different Hamiltonians that commute so that one of them can act as a quantum simulator is \tb{not easy and potentially} a vast limitation. However, in the following we will discuss two interesting cases and in particular show how to make use of the knowledge of $\hat h(t)$ in order to simulate infinite-range interactions with a system which exhibits only short-range nearest-neighbor interactions.

\begin{figure}[tb]
\centering
    \includegraphics[width=0.47\textwidth, height=0.3\textwidth]{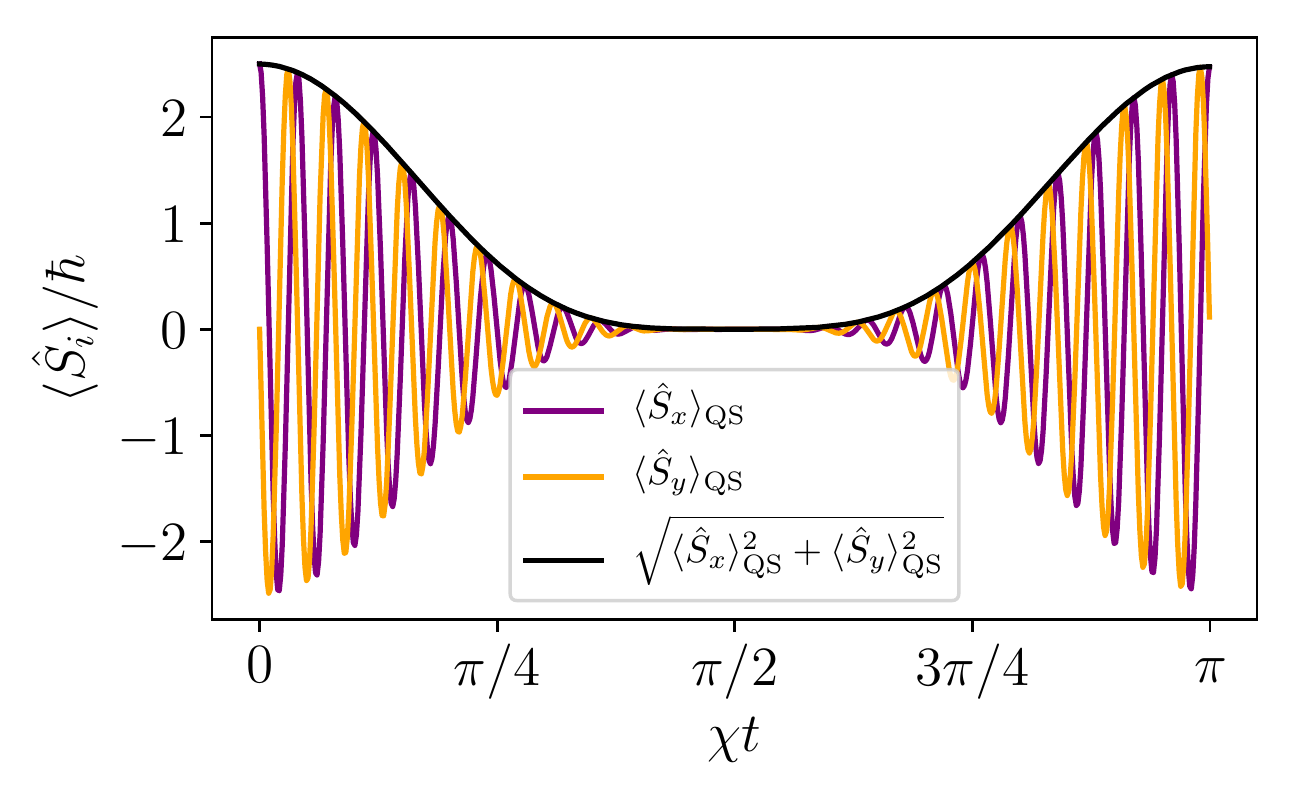}
 \caption{In order to calculate the time evolution of $\langle \hat S_x \rangle_{\mathrm{T}}$ in target system, it is necessary to measure how $\langle \hat S_x \rangle_{\mathrm{QS}}$ and $\langle \hat S_y \rangle_{\mathrm{QS}}$ depend on time in the quantum simulator system.  In the numerical simulations, we have set $\chi = 1$, $\beta/\alpha \approx 40.0$ ($\alpha \approx 1.299\sqrt{N-1} \sqrt{ \chi^2 + \chi \beta}$, see the main text for details), and $N = 5$ spins. }
    \label{fig:rep}
\end{figure}

\emph{Analog quantum simulators}---As a first example we will consider how to simulate the well-known one-axis twisting Hamiltonian \cite{PhysRevA.47.5138}
\begin{align}
	\hat H_{\mathrm{oat}} = \chi \sum_{i \leq j}^N \frac{\hat \sigma_i^z \hat \sigma_j^z}{4} = \chi \hat S_z^2,
\end{align}
where $\hat S_z = \sum_{i=1}^N \hat \sigma_i^z$ is the collective spin operator. Despite its simplicity, this Hamiltonian is known to generate a wide spectrum of many-body entangled states such as spin-squeezed, twin Fock, and Greenberger-Horne-Zeilinger states if the initial state is an eigenstate of the $\hat S_x$ operator with maximal eigenvalue, i.e., $\hat S_x|\psi \rangle = N/2|\psi\rangle$ \cite{PhysRevA.92.043622}. It can also be realized experimentally with ultra cold gases \cite{sorensen2001many} and trapped ions \cite{PhysRevLett.82.1835}. On the other hand, due to its formal simplicity, we can easily find a non-trivial and interesting Hamiltonian that commutes with the one-axis twisting Hamiltonian. It is straightforward to show that the Heisenberg XX model
\begin{align}
\label{eq:Hoat}
    \hat H_{\mathrm{XX}} = \frac{\beta}{4} \sum_{i=1}^{N-1}\left(\hat \sigma^x_{i}\hat \sigma^x_{i+1}  + \hat \sigma^y_{i}\hat \sigma^y_{i+1}\right),
\end{align}
commutes with $\hat H_{\mathrm{oat}}$, and therefore any eigenstate of $\hat H_{\mathrm{oat}}-\hat H_{\mathrm{XX}}$ will give the same dynamics under the action of the two different Hamiltonians. However, as $\hat H_{\mathrm{oat}}-\hat H_{\mathrm{XX}}$ \kgc{possesses a non-degenerate eigenspectrum}, such a simulator is fundamentally not very interesting \kgc{as it can only simulate the dynamics of eigenstates}. Nevertheless, one can add an arbitrary function of $\hat \sigma^z_i$ to the Heisenberg XX model and it will still commute with the $\hat H_{\mathrm{oat}}$ since $[\hat \sigma_i^z, \hat H_{\mathrm{oat}} ] = 0$. This then allows one to manipulate \tb{the form of}  $\hat h$ in such a way that the initial state $\hat S_x|\psi \rangle = N/2|\psi\rangle$ is also the eigenstate of $\hat h$, but not of the Hamiltonians building it. As an example we show that the Heisenberg XXX model with a staggered field
\begin{align}\label{eq:Hxxx}
    \hat H_{\mathrm{QS}} =& \frac{\beta}{4} \sum_{i=1}^{N-1}\sum_{j \in \{x,y,z\}}\hat \sigma^j_{i}\hat \sigma^j_{i+1}  
    + \frac{\alpha}2\sum_{i=1}^N (-1)^i\hat \sigma^z_{i},
\end{align}
can simulate one-axis twisting Hamiltonian in the limit $\beta \gg \alpha$ with $ \alpha = \sqrt{N-1} \sqrt{ \chi^2 + \chi \beta}$ and for an even number of spins, i.e, $N= 2k$ with $k \in \mathbb{N}$. Even though $\hat H_{\mathrm{QS}}$ and $\hat H_{\mathrm{oat}}$ are completely different, they realize the same dynamics. Most strikingly $\hat H_{\mathrm{QS}}$ contains only short-range nearest-neighbor interactions while $\hat H_{\mathrm{oat}}$ contains infinite-range interactions. Interestingly, we find that for an odd number of spins, $N = 2k+1$, and similar conditions, i.e, $\beta \gg \alpha$ and $ \alpha \approx 1.299\sqrt{N-1} \sqrt{ \chi^2 + \chi \beta}$, the Heisenberg XXX model with staggered field realizes both one-axis twisting and an effective rotation around $z$ axis with frequency given by $\alpha/N$. The rotation can be easily eliminated by moving to a frame which rotates around the $z$ axis with the same frequency but in the opposite direction, i.e., performing transformation $|\psi\rangle \rightarrow \hat U | \psi \rangle$ with $\hat U = \exp[i t (\alpha \hat S_z/N)]$ (note, however, that $\hat h$ does not have to be proportional to $\hat S_z$). This idea is similar to moving to a frame of reference rotating with the frequency of a pumping laser, which is a typical situation in quantum optics. We can therefore identify another interesting condition for a quantum simulator \kgc{using the connector}. \kgc{This is}, even if the initial state is not an eigenstate of $\hat h$ but $\hat h$ happens to trivially transform $|\psi\rangle$ (as in the case of a collective rotation or a translation), measuring an observable in the quantum simulator allows for \emph{measuring} it by performing a straightforward manipulation on the measured data, in this case given by
\begin{align}
	\langle \hat S_x(t) \rangle_{\mathrm{T}} = \sqrt{\langle \hat S_x(t) \rangle_{\mathrm{QS}}^2 + \langle \hat S_y(t) \rangle^2_{\mathrm{QS}}}.
\end{align}
The results of the numerical simulation and calculation of $\langle \hat S_x(t) \rangle_{\mathrm{T}}$ are presented in Fig.~\ref{fig:rep}.

\begin{figure*}[tb]
\centering
    \includegraphics[scale=0.59]{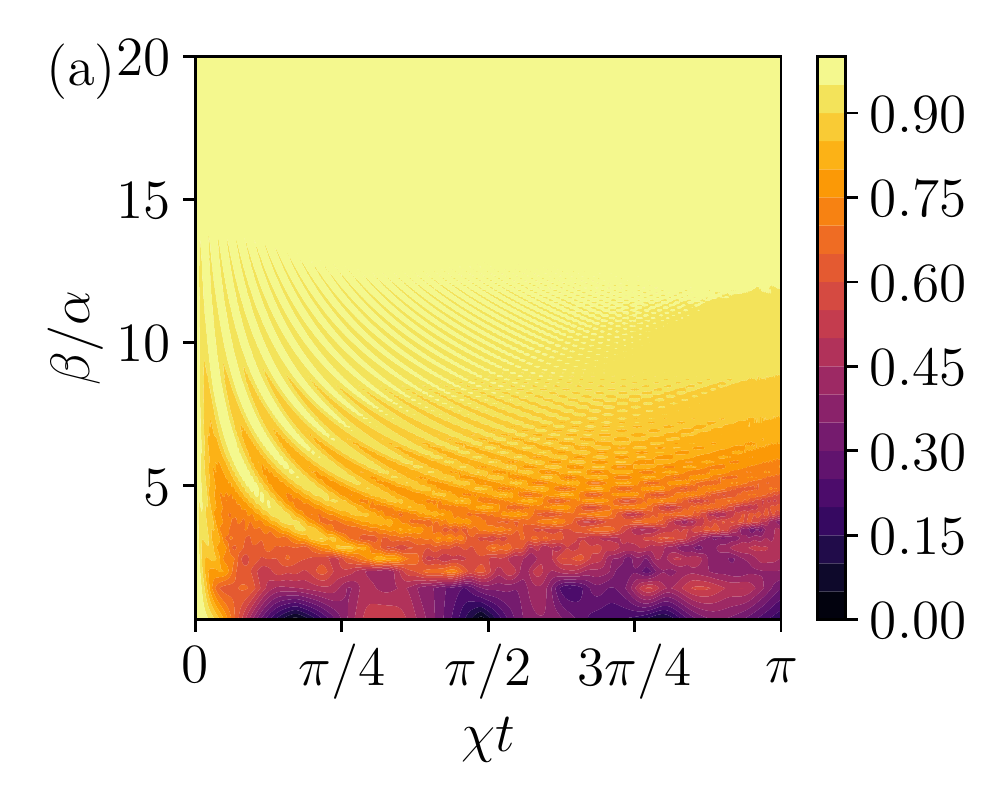}
    \includegraphics[scale=0.577]{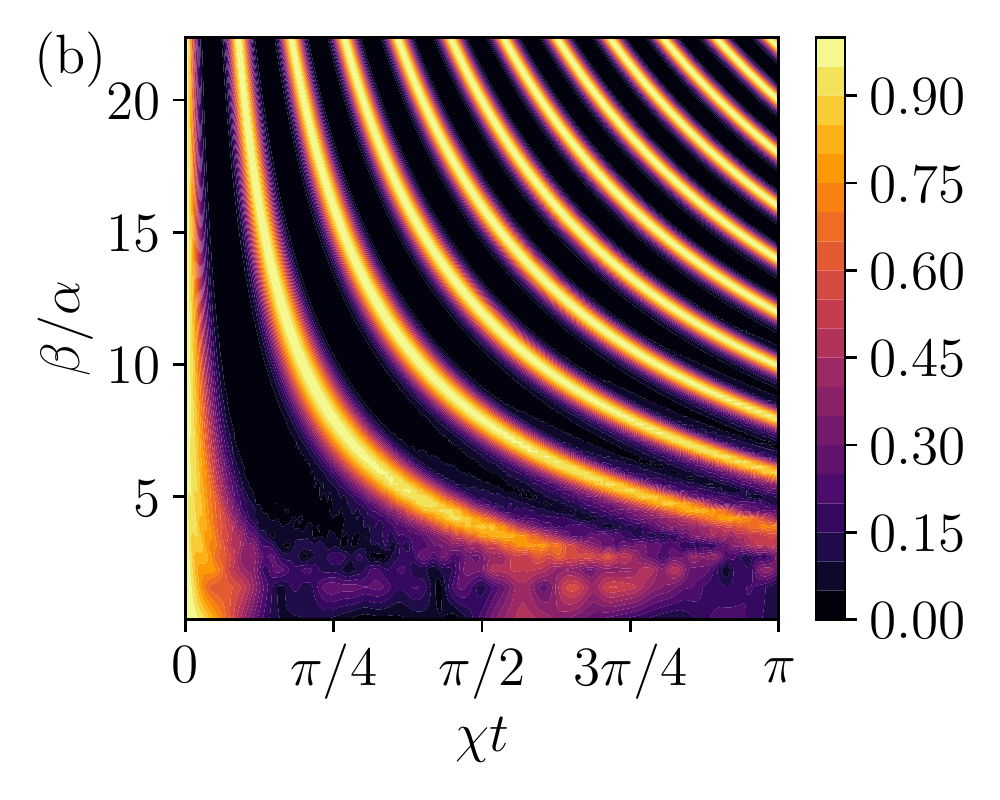}
    \includegraphics[scale=0.577]{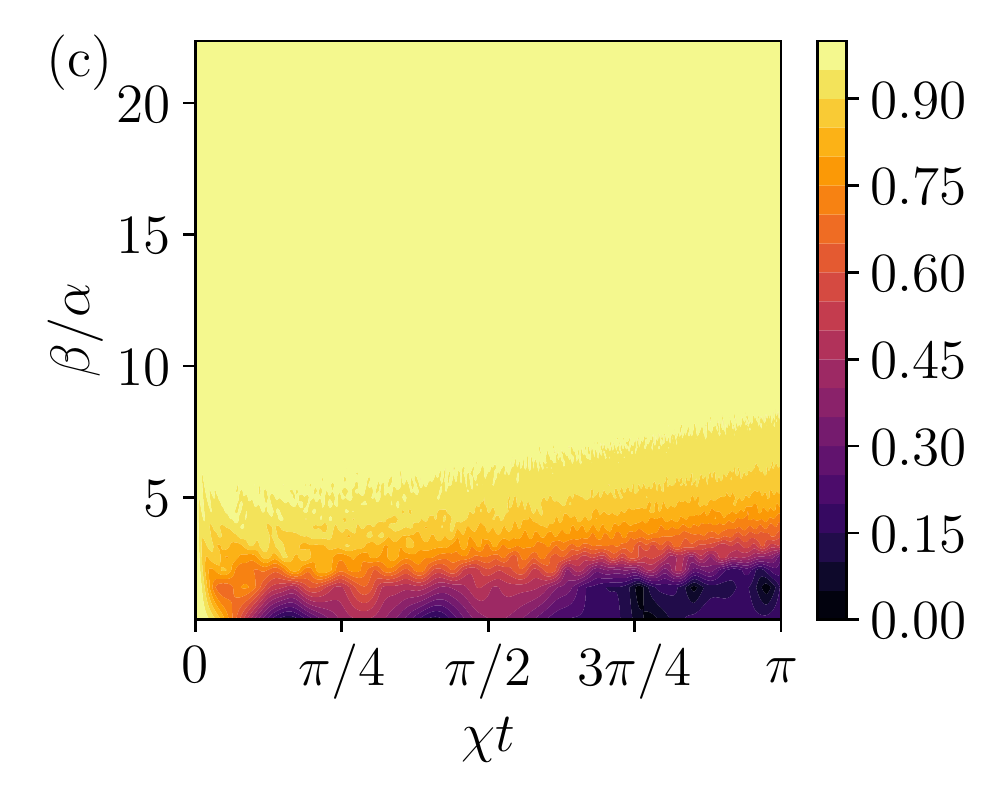}
 \caption{Fidelity between states generated by one-axis twisting [Eq.~(\ref{eq:Hoat})] and Heisenberg spin chain [Eq.~(\ref{eq:Hxxx})] as a function of time and $\beta/\alpha$ ($\alpha \sim \sqrt{\beta}$) for a system with (a) an even number of spins ($N = 6$) and (b) an odd number of spins ($N = 5$). In (c) the data for the odd number of spins is re-plotted using a frame of reference rotating with frequency $\omega = \alpha/N$ around the $z$-axis. In (b) one cannot only observe one-axis twisting but also rotation of the state (due to the lack of discrete translational symmetry), which can be removed by moving to a proper frame of reference. In the numerical simulations, we have set $\chi = 1$. Note that for $\chi t =\pi/2$ the state of the system is the maximally entangled Greenberger-Horne-Zeilinger state.}
    \label{fig:oatmim}
\end{figure*}


In order to \tb{investigate the robustness of} simulating the one-axis twisting dynamics with a Heisenberg XXX chain with a staggered field, we plot the fidelity between the states generated with these two Hamiltonians for two cases $N = 6$ and $N = 5$ as a function of time and $\beta/\alpha$ in Fig.~\ref{fig:oatmim}. \kgc{One can see why the condition given by Eq.~(\ref{eq:condgen}) does not require $\Im[{\xi(t)}]=0$. For times such that $\Im[{\xi(t)}]\ll 1$, the dynamics \kgc{governed by the simulator} still \tb{very much} resembles the dynamics governed by the target Hamiltonian.}

It can be also shown that Heisenberg XXX model with an arbitrary transverse field in the $z$ direction commutes with the special case of the Lipkin-Meshkov-Glick model $\hat H_{\mathrm{LMG}} = \hat S_x^2 + \hat S_y^2 +  \Omega \hat S_z$ \cite{PhysRevLett.99.050402} or with $\hat H = \sum_{n=1}^{\infty} \gamma_n\hat S_z^n$; and the Heisenberg XXX model without a transverse field commutes with a generalized two-axis-counter twisting Hamiltonian $\hat H_{\mathrm{tact}} = \chi(\hat S_x \hat S_y + \hat S_y \hat S_x) + \alpha \hat S_x + \beta \hat S_y + \gamma \hat S_z$.  However, the question of whether one can simulate non-trivial physics of these Hamiltonians using the connector approach remains open \tb{at this time}.
An interesting \tb{situation arises} when the two Hamiltonians do not commute. In such a case, the connector can be expressed as $\hat h = \sum_{n=1}^\infty t^n \hat A_n$, where $\hat A_n$ are operators that can be found according to the Baker-Campbell-Hausdorff formula. Also, when at least one of the Hamiltonians is time-dependent, it might lead to interesting possibilities of quantum simulation. All of these possibilities may relax constraints imposed on the universal analog quantum simulator, but we defer all of them to future investigations. Instead, we will focus now on the possibility of \kgc{using the connector operator} in the digital quantum simulator. 


\emph{Digital quantum simulator}---A digital quantum simulator \cite{lloyd1996universal} works by evolving a system forward using small and discrete time steps according to
\begin{align}\label{eq:trotter}
	e^{i \hat H_{\mathrm{T}}t} \approx  \left(e^{i \hat H_1 t/n}\ldots e^{i \hat H_{\ell}t/n} \right)^n.
\end{align}
By making $t/n$ small enough and using error correction protocols, this allows to simulate $\hat H_{\mathrm{T}}$ with an arbitrary precision.

This concept can also be applied to perform digital quantum simulation \kgc{using the connector operator}. If the time evolution interval is short enough, we can neglect the higher order commutators in Eq.~(\ref{eq:BCH}), i.e.,
\begin{align}\label{eq:BCHshort}
	\hat h(t) \approx \delta t\left(\hat H_{\mathrm{QS}} -\hat H_{\mathrm{T}} \right).
\end{align}
\kgc{In contrast to the situation where the Hamiltonians $\hat H_\mathrm{QS}$ and $\hat H_\mathrm{T}$ commute, here the eigenstates of $\hat h$ are different to the eigenstates of $\hat H_{\mathrm{QS}}$ and $\hat H_{\mathrm{T}}$. While this is in general a simplification, the price to be paid for it is that the eigenstates of $\hat H_\mathrm{QS}-\hat H_\mathrm{T}$ are only \tb{approximate} eigenstates of $\hat h$ for} short time intervals \tb{while} $\frac{i (\delta t)^2}{2}[\hat H_{\mathrm{QS}}, -\hat H_{\mathrm{T}}  ] \approx 0$. \tb{However, since during these} the two Hamiltonians will yield the same dynamics \tb{one can perform} stroboscopic dynamics \tb{by changing} $\hat H_\mathrm{QS}$ to $\hat H_\mathrm{QS}^\prime$ \tb{after every \emph{quantum kick}}. \tb{If the new Hamiltonian is chosen} such that the state after \tb{the last} \emph{quantum kick} is the eigenstate of the operator $\hat H_{\mathrm{QS}}^\prime -\hat H_{\mathrm{T}} $, one can then simulate $\hat H_T$ with the \kgc{\emph{quantum kicks} generated by $\{\hat H_{\mathrm{QS}},\hat H_{\mathrm{QS}}^\prime,\ldots, \hat H_{\mathrm{QS}}^{(n)}\}$, where $n$ labels the $n$th \emph{quantum kick}}. Naturally, the \emph{smaller} the commutator, the longer each \emph{quantum kick} can be applied for, and in the limit of the commutator going to 0, we recover the analog quantum simulator discussed in the previous section. In this sense, the analog quantum simulation is a special case of digital quantum simulation where the length of the \emph{quantum kick} can be infinitely long. 


Similarly as in the original idea of the digital quantum simulator, the digital quantum simulator using the connector operator has to be first accordingly prepared. In the former case, one has to use the so-called Trotter expansion, and in the latter case one has to ensure that $|\psi\rangle$ is an eigenstate of $\hat H_{\mathrm{QS}}^{(n)} -\hat H_{\mathrm{T}} $ after each \emph{quantum kick}. However, as the digital quantum simulator using the connector requires much fewer steps as the sequence of \emph{kicks} has to applied only once \kgc{instead of $n$ times} [see Eq.~(\ref{eq:trotter})]. The price to be paid for this simplicity in relation to the standard digital quantum simulator is the fact that for every initial state, one has to come up with a unique set of \emph{quantum kicks}. Nevertheless, given the fact that in the experiment only a tiny fraction of all possible quantum states can be addressed, it should not be viewed as a \kgc{major} obstacle (see Fig.~\ref{fig:scheme}). Also, depending on the particular \kgc{target Hamiltonian}, some quantum simulators will be better than others since some of them will minimize the commutator $[\hat H_{\mathrm{QS}}, -\hat H_{\mathrm{T}}]$ allowing thus for increasing the length of a single time step $\delta t$. 

Last but not least, one can think about combining the Trotter decomposition with the connector approach. Imagine that one has an operator $\hat O$ that commutes with the target Hamiltonian $\hat H_{\mathrm{T}}$ \kgc{or $\hat h$ can be easily calculated}. Then, as we have shown, for the \kgc{eigenstates} of $\hat O - \hat H_{\mathrm{T}}$, the unitary evolution operators $\exp({-i t \hat O})$ and $\exp({-i t \hat H_{\mathrm{T}}})$ will yield the same dynamics. As a consequence, if $\hat O$ is much simpler than $\hat H_{\mathrm{T}}$, decomposing $\exp({-i t \hat O})$ should become much easier than decomposing $\exp({-i t \hat H_{\mathrm{T}}})$.


 
\emph{Conclusions and outlook}---By using the knowledge of a connector operator \tb{of} two Hamiltonians residing in the same Hilbert space, we have proposed a way of simulating the dynamics governed by one Hamiltonian using a different one. As an example of an analog quantum simulation, we have shown how to implement the one-axis twisting Hamiltonian in the Heisenberg XXX model with a staggered field. Using the connector, we have also proposed an alternative approach to digital quantum simulators. Instead of trying to build the target Hamiltonian $\hat H_\mathrm{T}$ out of many small steps, one has to apply short \emph{quantum kicks} with a quantum simulator Hamiltonian $\hat H_{\mathrm{QS}}$ such that after each \emph{quantum kick} the state is an eigenstate of the $\hat H_{\mathrm{QS}}-\hat H_{\mathrm{T}}$ operator. This can significantly reduce the complexity of a digital quantum simulator. The price being paid is the fact that not all initial states can be \emph{easily} used in the simulator (see Fig.~\ref{fig:scheme}). However, given the fact that not all initial states can be prepared in an experiment, by appropriately tuning the parameters of the simulator one should be able to simulate non-trivial physics of other systems. We have also identified interesting possibilities for future research including analog quantum simulation in the case when two Hamiltonians, $\hat H_{\mathrm{QS}}$ and $\hat H_{\mathrm{T}}$, do not commute or when the target Hamiltonian is time-dependent. A fascinating question that remains to be addressed in future research is whether the presented framework can be used with dissipative time evolution.

The results presented in this work might have direct implications in many branches of modern physics as well as quantum chemistry \cite{kassal2011simulating,arguello2019analogue,RevModPhys.92.015003} and quantum biology \cite{lambert2013quantum,davies2004does}, and can be tested in most of the current quantum simulator experimental set ups. However, the most striking consequences pave a way towards an approach to simulating dynamics not only with other systems but with other Hamiltonians. This might relax the constrains on the universal quantum simulator as it is not necessary to use exactly the same Hamiltonian to simulate the physics of some other Hamiltonians. On the downside, even though in certain situations it might be easier to perform quantum simulations \kgc{exploiting the connector operator}, in general it might be more challenging to find proper quantum simulators allowing for taking \kgc{advantage of this framework of connector}. 

Additionally, we have proposed a method for creating many-body entangled states, including the spin-squeezed and the maximally entangled Greenberger-Horne-Zeilinger state, in a system exhibiting exclusively nearest-neighbor interactions. \kgc{This might become extremely useful for the quantum computer architectures based on superconducting qubits as they typically exhibit only nearest or next-nearest neighbor interactions \cite{kjaergaard2020superconducting}}


\begin{acknowledgments}
\emph{Acknowledgements}---Simulations were performed using the open-source QuantumOptics.jl framework in Julia~\cite{kramer2018quantumoptics}. K.G. would like to acknowledge discussions with Tomasz Maci\c a\.zek, Mohamed Boubakour, Friederike Metz, Lewis Ruks, Hiroki Takahashi, and Jan Ko{\l}ody\'nski. This work was supported by the Okinawa Institute of Science and Technology Graduate University. K.G. acknowledge support from the Japanese Society for the Promotion of Science (JSPS) grant number P19792. A.U. acknowledges a Research Fellowship of JSPS for Young Scientists. K.G. would like to thank Linda Aleksandra Gietka for inspiration, Simon Hellemans for his support, and Micha{\l} Jachura for reading the manuscript.
\end{acknowledgments}

\end{document}